\documentclass[aps,prl,twocolumn,showpacs,superscriptaddress,groupedaddress]{revtex4}
\usepackage{graphicx}  
\usepackage{dcolumn}   
\usepackage{bm}        
\usepackage{amssymb}   

\newcommand{\be}{\begin{equation}}
\newcommand{\ee}{\end{equation}}
\newcommand{\bea}{\begin{eqnarray}}
\newcommand{\eea}{\end{eqnarray}}

\usepackage{setspace}
\usepackage{amsmath}
\usepackage{amssymb}
\usepackage{fancybox}
\usepackage{listings}
\usepackage{url}

\begin{document}

\title{Fractional Helical Liquids and Non-Abelian Anyons in Quantum Wires}
\author{Yuval Oreg}
\affiliation{Department of Condensed Matter Physics, Weizmann Institute
of Science, Rehovot, 76100, Israel}
\author{Eran Sela}
\affiliation{ Raymond and Beverly Sackler School of Physics and Astronomy, Tel-Aviv University, Tel Aviv, 69978, Israel}
\author{Ady Stern}
\affiliation{Department of Condensed Matter Physics, Weizmann Institute
of Science, Rehovot, 76100, Israel}

\begin{abstract}
We study one dimensional wires with spin-orbit coupling. We show that in the presence of Zeeman field and strong electron-electron interaction a clean wire may form \emph{fractional helical liquid} states with phenomenology similar to fractional quantum Hall liquids.  Most notably, the wire's two terminal conductance is predicted to show fractional quantized conductance plateaus  at low electron density. When the system is proximity-coupled to a superconductor, fractional Majorana bound states may be stabilized. We discuss how disorder destabilizes these fractional phases. Possible experimental realizations of similar states in double wire systems are discussed.

\end{abstract}


\maketitle

\emph{Introduction:}
Two-dimensional topological insulators have one dimensional helical modes at their edges. In these edges counter-propagating modes are related by time reversal \cite{Bernevig06,Konig07}. These systems are most easily understood when viewed as integer-quantum-spin-Hall systems, in which electrons with opposite spins are at integer quantum Hall states with filling $\nu=\pm 1$. When properly subjected to proximity coupling to superconductors and ferromagnets these edges may host localized zero energy Majorana modes \cite{Fu09}.

Similar phenomenology occurs in one-dimensional (1D) wires where electrons are subjected to spin-orbit coupling and an external Zeeman field~\cite{Oreg10}. The two-terminal conductance of such a wire varies non-monotonically as a function of increasing  chemical potential, changing  from zero (for a fully depleted wire), to a $2e^2/h$ plateau, then to a $e^2/h$ plateau at which the wire is helical, and back to $2e^2/h$ (see Fig.~1) as observed in a recent experiment~\cite{dgg10}. Furthermore, it was shown theoretically \cite{Oreg10,Lutchyn10} and experimentally~\cite{Mourik12,Das12,Deng12} that in proximity to a superconductor  the helical state gives rise to end modes that are Majorana modes.

In two-dimensions (2D) interactions between electrons may give rise to fractionalized phases, whose edges are gapless Luttinger liquids.
In this work we study how this fractionalization may be reflected in quantum wires. Considering the  same setting used for the one-dimensional helical quantum liquids in \cite{Oreg10,Lutchyn10}, namely a single-mode wire where electrons are subjected to spin-orbit coupling and a Zeeman field, we examine the conditions under which one-dimensional ``fractional helical liquids'' may be formed, their properties and their sensitivity to disorder. The crucial dimensionless number for our analysis is the ratio of the wire's Fermi momentum $k_{\rm F}$ to the momentum characterizing spin-orbit coupling, $k_{\rm SO}$. We find that when this ratio is an odd integer $2n+1$ a fractional helical liquid state may form ($n$ is a non negative integer). In such a liquid the two-terminal conductance of the wire is $\frac{e^2}{h}\frac{2}{1+(2n+1)^2}$. The non-monotonic dependence of the conductance on chemical potential becomes then much richer.

\begin{figure}[h]
\begin{center}
\includegraphics*[width=\columnwidth]{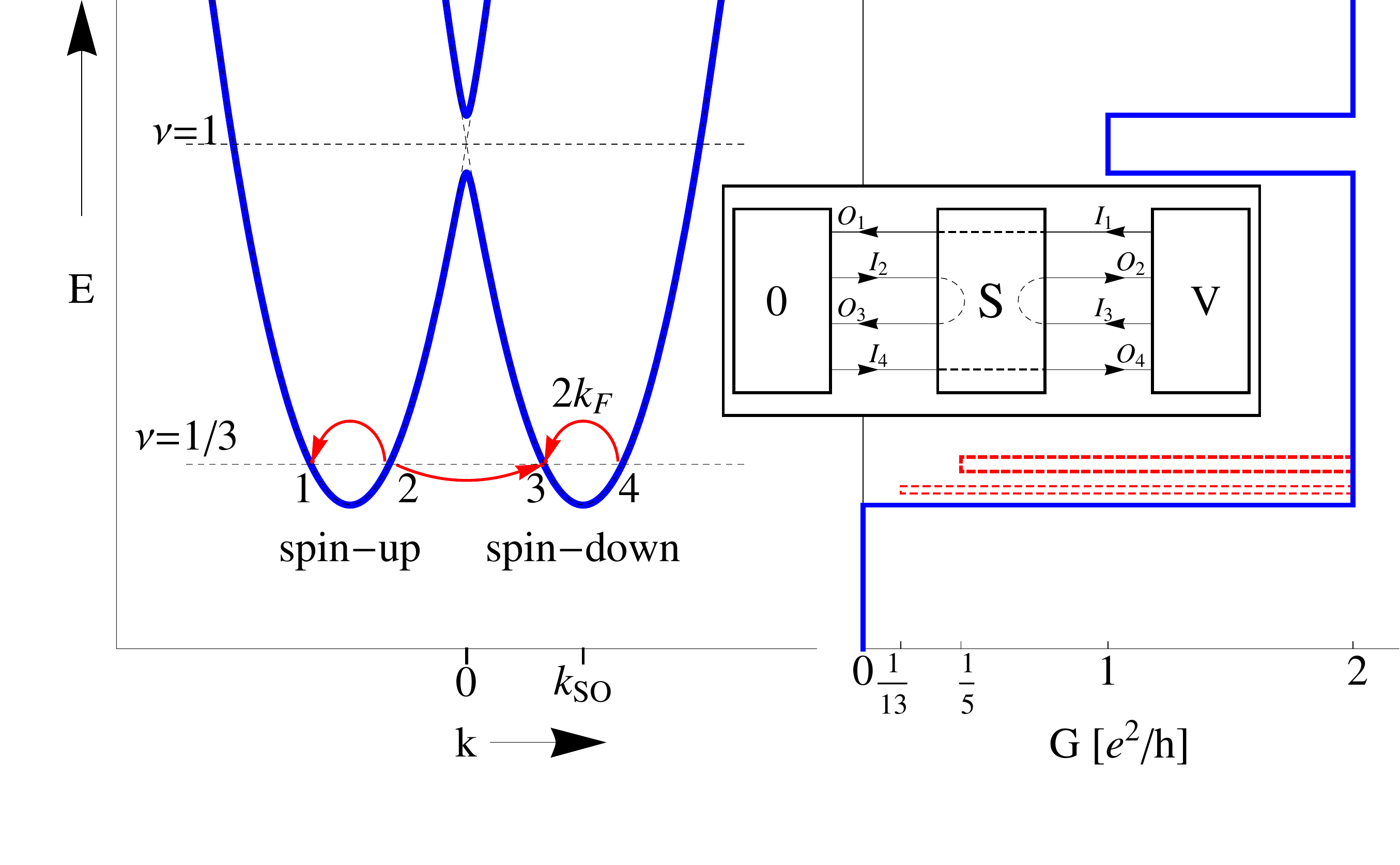}
\vspace{-1cm}
\caption{\label{fg:1} Left panel: spin-orbit split bands in a quantum wire, and a small energy gap opening due to magnetic field near the band-crossing point. Right panel: a sketch of the zero temperature conductance dI/dV versus gate voltage. Inset: two probe measurement setup where the fractional state is stabilized in the central region between two non-interacting leads; dashed lines represent the scattering pattern of bosons for $n=0$.
}
\end{center}
\end{figure}

 Furthermore, in 2D counter-propagating edge modes
 may occur in fractionalized phases as well. Examples range from fractional topological insulators, that are presently theoretical, to graphene electron-hole bi-layers, which are experimentally feasible. It was shown theoretically that in proximity to superconductors and ferromagnets the counter-propagating edge modes may be gapped to host localized fractionalized Majorana modes, that are non-abelian anyons of a new type~\cite{Clarke12,Lindner12,Cheng12,Vaezi13}.  We show here that proximity-coupling of a 1D fractional helical liquid to a superconductor  gaps its spectrum, and gives rise to fractionalized Majorana modes akin to those found in~\cite{Clarke12,Lindner12,Cheng12,Vaezi13}. The resulting gapped state is not within the list of possible gapped one-dimensional phases classified in Refs.~\onlinecite{Fidkowski11,Turner11}.
 It then comes as no surprise that these states are not stable to disorder.

Our analysis benefits greatly from the study of the two-dimensional quantum Hall effect in terms of a set of one-dimensional coupled quantum wires. In this approach, carried out in \cite{Kane02} and \cite{Teo11}, the 2D plane is seen as made of an array of coupled wires. The spectrum of each wire is parabolic, and the parabolas are shifted with respect to one another in momentum space by the magnetic field. The number of wires is proportional to the sample's width, and approaches infinity in the thermodynamic limit. As we show below, our system corresponds to the system considered in Ref.~\onlinecite{Kane02} when the number of wires is two.

While the main focus in this manuscript is on a single wire with spin-orbit coupling that shifts the free electron spectrum parabolas, a similar effect may occur in a pair of parallel wires subjected to a magnetic field perpendicular to their plane. In these wires, spin polarization by the magnetic field and a proper tuning of the Fermi level in each wire may stabilize fractional helical phases. Such wires were experimentally realized by means of cleaved edge overgrowth~\cite{Auslaender02}.

\emph{Model:} Our model of interest consists of a spinfull, single band wire, parallel to the $\hat{x}$ direction, with Rashba spin-orbit (SO) coupling $u$, and with a magnetic field $B$ nonparellel to the SO quantization axis. The Hamiltonian is
\begin{equation}
\label{eq:H}
 {\cal H} = \frac{k^2}{2 m}-\mu + u \hat \sigma^y k +B \hat \sigma^z,
 \end{equation}
 with $\hat \sigma^i,\;  i=x,y,z $ being the Pauli matrices (here and henceforth we work with a system of units in which $\hbar=1$), and $k$ being the momentum.
 The SO coupling causes a shift of the momenta $k$ of the free electrons parabolic spectrum $E(k)$. The spin up electronic dispersion (up with respect to a direction $\hat y$ set by the SO coupling) is shifted to the left by a momentum $k_{\rm SO}=m u$ while spin down is shifted to the right by $k_{\rm SO}$ (see Fig.~1). The two parabolas intersect at zero momentum so that an application of uniform (Zeeman) magnetic field (not parallel  to $\hat y$) opens a gap in the spectrum near zero momentum. Two electron helical modes, one propagating to the left and the other to the right, remain free.

 Away from the gap there are four Fermi points [with wave vector $ k_l= \pm (k_{\rm SO}\pm k_F)$] and therefore we define $4$ fermionic fields $\psi_l$ with $l=1,2$  corresponding  to the left and right moving modes of $\sigma_y =1$ and $l=3,4$ corresponding to the left and right moving modes of $\sigma_y =-1$ (see Fig. \ref{fg:1}).

 Interactions affect this Hamiltonian in two ways. Small momentum interactions are accounted for by introducing  Luttinger liquid parameters, as usual. Large momentum interactions need to be elaborated on for our discussion. As pointed out by Kane {\it et al.} in Ref. \onlinecite{Kane02} in the context of the fractional quantum Hall (FQH) effect, multi-electron processes that involve large momentum transfer may lead to an opening of an energy gap at proper values of the chemical potential. Following~\cite{Kane02} we focus on the interaction term
 \bea
\label{eq:O}
\mathcal{O}_n^B =g_B \left({\psi_{1}^\dagger} {\psi_{2}^{\phantom \dagger}}\right)^n
\psi_2 \psi_3^\dagger \left({\psi_{3}^\dagger} {\psi_{4}^{\phantom \dagger}}\right)^n.
\eea
 For $n=0$ this term reduces to the mixing of modes 2  and 3 at the band crossing point, $\mathcal{O}_0^B=g_B \psi_2 \psi_3^\dagger$. For $n=1$ this process takes 3 particles across the 4 Fermi points, $2 \to 1$, $2 \to 3$, $4 \to 3$; see arrows in Fig.~(\ref{fg:1}). Such a term (together with its hermitian conjugate) is odd under time reversal, hence it requires magnetic field to be present. It is generated at second order in the bare interaction strength  $g_B \propto B U_{2 k_{\rm F}}^2$, with $U_q$ the interaction potential.

 The combined Hamiltonian is most conveniently treated by bosonization. The four chiral bosonic fields corresponding to the four fermionic fields are $\psi_l=e^{i \phi_l};\;\; \psi^\dagger_l \psi^{\phantom \dagger}_l= (-1)^l \frac{1}{2\pi} \partial_x \phi_l$. The bosonic Lagrangian density is
 \begin{eqnarray}
 {\cal L}&=&-\frac{1}{4\pi}\sum_l(-1)^l\partial_t\phi_l\partial_x\phi_l-\sum_{lj}\partial_x\phi_l U_{lj}\partial_x\phi_j  \nonumber \\
 &+&g_B \cos{\left\{\phi_3-\phi_2+ n\left[(\phi_3-\phi_4)-(\phi_2-\phi_1)\right]+k x\right\}}\nonumber\\
 &-& \frac{1}{2\pi}E\sum_l(-1)^l\phi_l.
 \label{eq:originalL}
 \end{eqnarray}
The positive definite symmetric matrix $U$, describing the bare velocities of the fields and the low momentum transfer interaction, is not universal, here $k= k_3-k_2 +n [(k_3-k_4)-(k_2-k_1)]$. When $k \approx 0$, the term proportional to  $g_B$ gaps out two of the four modes in the spectrum.
Using the parabolic spectrum when $B \ll \Delta_{SO}=m u^2/2$ we find that $k=0$ for $k_{\rm F} = k_{\rm SO}/(2n+1)$. With the relation $\rho=2 k_F/\pi$, where $\rho$ is the density of the spinfull wire, we find that a gap will open for filling $\nu= \frac{k_F}{k_{SO}}= \frac{\pi \rho}{2 mu}=\frac{1}{2n+1}$.  Notice that a component of the Zeeman field in the SO coupling direction ($\hat y$) will change the conditions for $k=0$, the generalization is straightforward.

The $g_B$ term in the Lagrangian (\ref{eq:originalL}) motivates us to introduce a new set of chiral bosonic fields, denoted by $\eta_l=\sum_j M_{lj}\phi_j$, given explicitly by:
 \begin{equation}
 \left(\begin{array}{cc}
   \eta_1  \\
   \eta_2 \\
   \eta_3\\
   \eta_4
 \end{array} \right)
 =
  \left(\begin{array}{cccc}
   n+1 &-n   &0&0\\
   -n  & n+1 &0&0\\
   0   &0    & n+1 &-n \\
   0   &0    & -n & n+1
   \end{array} \right)
   \left(\begin{array}{cc}
   \phi_1  \\
   \phi_2 \\
   \phi_3\\
    \phi_4
 \end{array} \right).
  \label{eq:transformation}
 \end{equation}

 Notice that for $n=0$ we have $\eta_l=\phi_l$.  In terms of the $\eta$ fields, the Lagrangian density becomes,
 \begin{eqnarray}
 {\cal L}&=&-\frac{1}{4\pi}\sum_l\frac{(-1)^l}{2n+1}\partial_t\eta_l\partial_x\eta_l
   -\sum_{lj}\partial_x\eta_l {\tilde U}_{lj}\partial_x\eta_j \nonumber \\
   &+& g_B\cos\left(\eta_3-\eta_2\right) - \frac{1}{2\pi}E\sum_l\frac{(-1)^l}{2n+1}\eta_l,
 \label{transformedL}
 \end{eqnarray}
with $\tilde U = (M^{-1})^T U M^{-1}$.

For $B \gg \Delta_{\rm SO}$ the description in terms of four bosonic fields ceases to be valid (as $k_2$ and $k_3$ do not exist anymore). However, similar to the case for $n=0$ when the system is fully gapped (e.g. by proximity to a superconductor) the gap is not necessarily  closed when $B \gtrsim  \Delta_{\rm SO}$ and a phase transition would not necessarily occur then.

We now turn to discuss several properties of the wire that follow from the analysis of Lagrangian (\ref{transformedL}) in the limit of large $g_B$.

\emph{Two terminal conductance:}
In a two terminal conductance measurement one connects two noninteracting reservoirs to the wire. We model this  situation by a central interacting section connected to the reservoirs via two non interacting sections. By assumption, the transition between the sections is smooth on the scale of $k_{\rm F}$ so that a $2 k_{\rm F}$ scattering does not occur. The DC conductance can then be found by studying a scattering problem in which electron currents incident from the reservoirs scatter in the central region as shown in the inset to Fig.~\ref{fg:1}.

  To solve the scattering problem it is useful to define incoming and outgoing current vectors. The incoming currents are $\vec I  =\left(I_1,I_2, I_3,I_4\right)^T$, with $I_l = \frac{e}{2 \pi} \partial_t \phi_l|_{x=x_l}$ and $l=1,\dots,4 $. The points $x_l$ are at $+\infty$ for the left movers $l=1,3$ and $-\infty$ for the right movers $l=2,4$ (see Fig. \ref{fg:1}). A similar definition holds for the outgoing current $\vec O  =\left(O_1,O_2, O_3,O_4\right)^T$, with $x_l\rightarrow -x_l$. Assuming now that incoming left movers (1,3) both emanate from a reservoir at potential $V$ and the right movers (2,4) from a reservoir at potential zero, we have $\vec I = \frac{e^2}{h} V \left(1,0,1,0\right)^T$, and $\vec O=S \vec I$, with $S$ being the scattering matrix. The conductance $G$ is then given by:
  \begin{eqnarray}
  G V &=&I_{1}+ I_{3}-O_{2}-O_{4} =- I_{2}- I_{4}+O_{3}+O_{1} \nonumber \\
  G/(e^2/h)   &=&2- (0,1,0,1)S(1,0,1,0)^T \nonumber \\
              &=& (1,0,1,0)S(1,0,1,0)^T.
      \label{eq:GS}
  \end{eqnarray}
Next we should find the scattering matrix $S$. In the free case $g_B=0$ and the scattering matrix is equal to the  $4\times 4$ identity matrix. Simple algebra then gives: $G=2 e^2/h$. For the $n=0$ case with $g_B \rightarrow \infty$ the chiral field $\phi_2$ is fully reflected to $\phi_3$ (see inset to Fig.~\ref{fg:1}), and this leads to a scattering matrix $S_0$ equal to a matrix in which the second and the third rows of the rank $4$ identity matrix are interchanged. For that case  $G = e^2/h$ as expected.

Determination of the scattering matrix for $n\ne 0$ follows directly from the structure of the Lagrangian Eq.~(\ref{transformedL}). For $g_B \rightarrow \infty$ the cosine term implies that $\eta_2(x) = \eta_3(x)$ for $-L/2<x<L/2$ which is the region of the wire where the electrons interact. Using this equation at $x=-L/2$ and at $x=+L/2$ gives two relations
\bea
\label{bc1}
(n+1) I_2-n O_1&=& (n+1) O_3 - n I_4, \nonumber \\
(n+1) O_2-n I_1&=& (n+1) I_3 - n O_4.
\eea
A second pair of equations is obtained by realizing that fields $\eta_1$ and $\eta_4$ propagate freely through the interacting wire implying $\eta_{1,4}(-L/2)=\eta_{1,4}(L/2)$. In terms of the original fields the equations for $\eta_1$ and  $\eta_4$ become
\bea
\label{bc2}
(n+1)O_{1} - n O_2 &=&(n+1) I_1- n O_2, \nonumber \\
(n+1)I_{4} - n O_3 &=&(n+1) O_4 - n I_3,
\eea
respectively. This set of four equations can be written in a matrix form as $ A \vec{O}=B \vec{I}$ where $A,B$ are $4 \times 4$ matrices, giving the desired result $S = A^{-1} B$. Substituting this result for $S$ in (\ref{eq:GS}) we find the two terminal conductance:
  \begin{equation}
  \label{eq:G2}
  G = \frac{e^2}{h} \frac{2}{1+(2n+1)^2}  = \frac{e^2}{h} \frac{2 \nu^2}{\nu^2+1}.
  \end{equation}
For $n=0$ we find $G=\frac{e^2}{h}$ as expected, for $n=1$ ($\nu=1/3$) $G=\frac{1}{5}\frac{e^2}{h}$ and so on. In general, unlike the FQH effect the conductance is not equal to the filling factor $\nu$. Similar to the two modes on the edges of a FQH 2D system, in the present case two modes remain untouched by the massive term that mixes $\eta_2$ and $\eta_3$. However, unlike the 2D system were the free modes on different edges emanate from different reservoirs and propagate freely, in the present case the free propagating modes in the interacting section are mixed at the boundary between the interacting and the noninteracting sections. The physical proximity of the two counter-propagating modes makes them susceptible to disorder, as we will discuss below. We note that as in the case of a Luttinger liquid describing quantum Hall effect or quantum wires, the low momentum interactions $U_{lj}$ do not affect the two terminal conductance~\cite{Oreg95,Maslov95,Safi95,Ponomarenko95,Oreg96a}.

Refs. \onlinecite{Kane02, Teo11} show that in the limit of an infinite number of spinless wires whose parabolic dispersion is shifted by a magnetic field the two-terminal conductance approaches the expected value of the quantized Hall conductivity. The case we study corresponds to two wires. In the appendix we study numerically the case of $N$ wires, and show that the deviation of the two terminal conductance from the quantized Hall conductivity decreases very quickly with $N$, and is the order of $e^{-N}$ for $N\ge 5$.
This reflects the fact that the tunneling between the edge modes is exponentially small. All the values that we find for the two terminal conductance of $N$ wires at filling $\nu=k_{\rm F}/k_{\rm SO}= 1/(2 n+1)$ conform to the expression
\bea
\label{eq:GN}
\frac{G_N}{{e^2}/{h}}=\nu \frac{(\nu+1)^N-(\nu-1)^N}{(\nu+1)^N+(\nu-1)^N}.
\eea

\emph{Measurements of individual elements of the $S$-matrix:}  The individual elements of the $S$-matrix may be measured in a setup such as used by Auslaender \emph{et. al}~\cite{Auslaender02}. In this setup two parallel wires are subjected to a magnetic field perpendicular to their plane. If the spins are polarized and tunneling between the wires is enabled, this system maps onto the system we study here, with the wire index playing a role of a pseudospin. The two wires are connected to four  separate reservoirs, allowing a separate control of the incoming currents $I_1 \dots I_4$. For example, the application of  a voltage difference $V$ between the wires corresponds to an incoming current $\vec{I}=(V,V,0,0)$, and results in a net current flowing between the wires. This current is $I_s = G_s V = (0,0,1,1 )S(V,V,0,0)$. Using the above $S$ matrix this current is found to be equal to the charge current that flows when voltage is applied between the right and left moving modes, i.e., $G_s=G$. Interestingly, even when the current from wire $1$ emanates equally from the two reservoirs the wire is connected to, the magnetic field between the wires results in a current partitioning in wire $2$ that is not symmetric between the left and right drains. Rather, the difference between the currents flowing to the left and right in wire $2$ is  $ (0,0,1,-1 )S(V,V,0,0)=\frac{e^2}{h} \frac{2(2n+1)}{1+(2n+1)^2}V$.

\emph{Disorder:} Disorder that leads to scattering between the free left and right movers spoils the quantized conductance. This is obvious in the $n=0$ case, where interactions are absent. For $n\ne 0$ single electron tunneling is of the form $\psi^\dagger_l \psi_{l'}$. In the
 bosonized form it is $\lambda_{ll'} \cos(\phi_l-\phi_{l'})$. To find their relevance we write them in terms of the $\eta$'s. Since $\eta_2$ is pinned to $\eta_3$ by the $g_B$ term,  $\eta_2+\eta_3$ fluctuates wildly.  As a consequence, a term that contains $\eta_2+\eta_3$ is irrelevant, while the combination $\eta_2-\eta_3$ can be approximated as a constant. These considerations make the $\lambda_{14}$ and $\lambda_{23}$ the only relevant single particle terms. These terms translate to  $\cos{\left [\alpha(\eta_1-\eta_4)+\beta(\eta_2-\eta_3)\right ]}$, where $(\alpha,\beta)=\frac{1}{2n+1}(n+1,n)$ for $\lambda_{14}$ and $(\alpha,\beta)=\frac{1}{2n+1}(n,n+1)$ for $\lambda_{23}$. Both terms scatter charge between the two compressible modes $\eta_1$ and $\eta_4$, and affect the two terminal conductance.

 Generally, the scaling dimensions for these two terms makes them relevant. For repulsive interactions, these dimensions are bounded from above by $2\alpha^2/(1+\nu^2)$, see Eq.~(\ref{scalingD}) in the appendix. However, the divergences associated with the relevance of impurities are cut-off here by the finiteness of the wire, as well as by finite voltage and temperature.

 Multi-particle scattering by impurities may be even more relevant than single particle scattering. For example, $4k_F$ scattering of the form $\cos{(\phi_1-\phi_2+\phi_3-\phi_4)}=\cos{\nu(\eta_1-\eta_4-\eta_2+\eta_3)}$ leads to a scaling dimension bounded by $2\nu^2/(1+\nu^2)$. This backscattering operator corresponds to a quasiparticle tunneling in the FQH 2D limit.

\emph{Fractional charges:} In the $n=0$ case a soliton (i.e. a kink between $0, 2\pi$ in the $\eta_2-\eta_3$ field) describes an electron or hole with charge $C=\pm e$. Since $\sum_l (-1)^l \phi_l=\frac{1}{2n+1} \sum_l (-1)^l \eta_l$ it is readily understood that for general $n$ a kink in the the $\eta$ fields carries a charge \cite{Kane02}  $$C=\pm \frac{e}{2n+1}=\pm \nu e.$$ We note, however, that in 1D those quantized charges can not be isolated since the free modes (composed of $\eta_1$ and $\eta_4$) will screen those charges in a way that depends on the low momentum interaction $U_{lj}$.

\emph{Fractional Majorana modes:}
So far we considered a compressible helical liquid, in which the fractionalized nature of the wire reflects itself in the fractional two terminal conductance. The interaction term  $g_B\cos(\eta_3 - \eta_2)$ gaps two of the four gapless modes in the wire, but leaves the other two gapless. By coupling the wire to a super-conductor through the proximity effect, the wire may be fully gapped. In the $n=0$ case a coupling to superconductor leads to the formation of Majorana end modes~\cite{Oreg10,Lutchyn10}. We will show now that in the $n \ge 1$ case proximity to  a superconductor leads to the formation of fractional Majorana modes.

\begin{figure}[h]
\begin{center}
\includegraphics*[width=80mm]{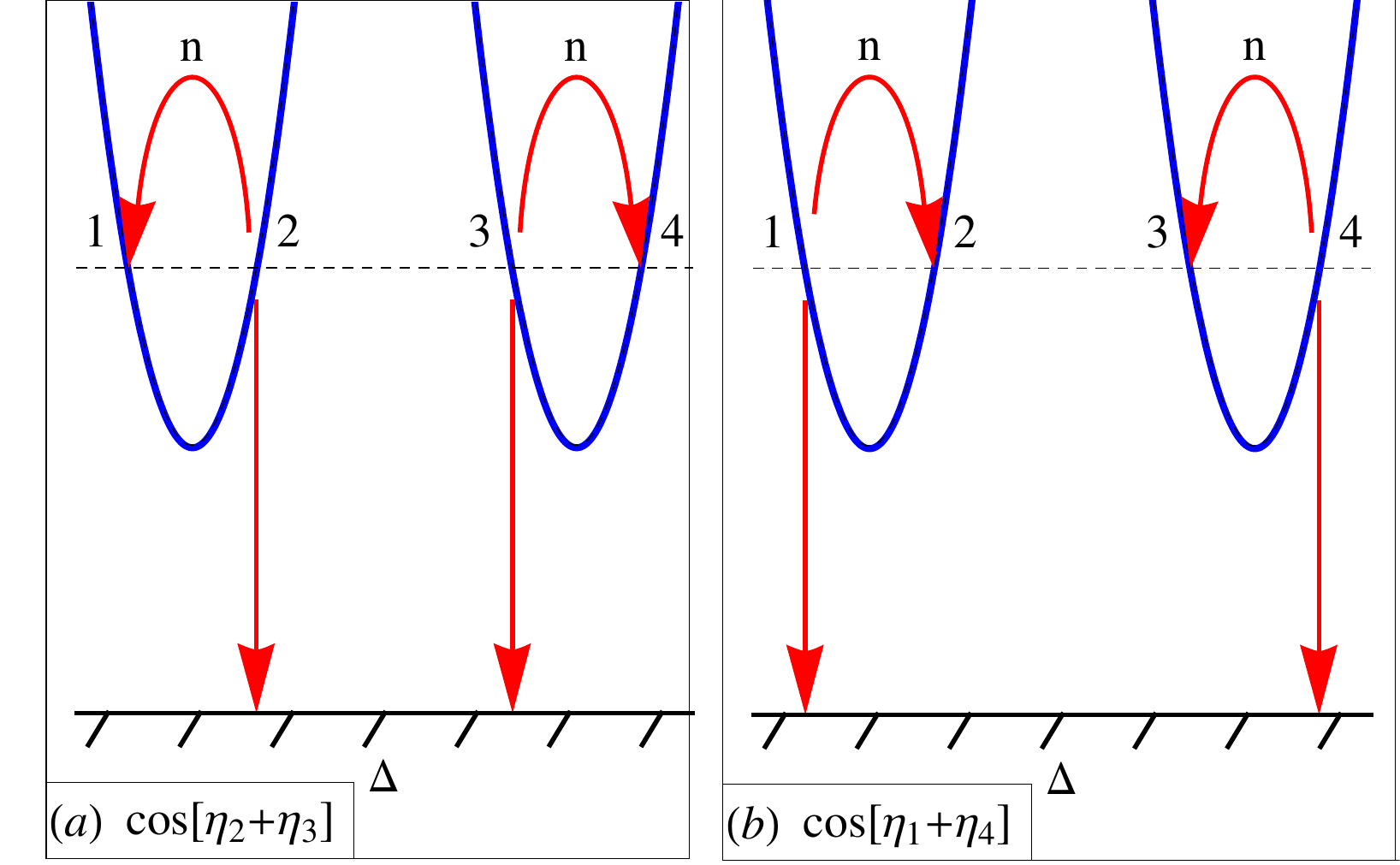}
\caption{\label{fg:3} (a) Superconducting analog of the process in Fig. 1. (b) Many particle process inducing a superconductivity gap in the fractional helical liquid.
}
\end{center}
\end{figure}

The proximity induced superconducting  mean field Hamiltonian projected to the four Fermi points is $ \mathcal{H}_{\Delta} = \Delta \int dx (\psi_1(x) \psi_4(x) + \psi_2(x) \psi_3(x) +{\rm{H.c.}})$. Bosonizing, we find $ \mathcal{H}_{\Delta} \sim \Delta \cos(\phi_1 + \phi_4) + \Delta \cos(\phi_2 + \phi_3)$. Both operators are irrelevant for $n\ge 1$ since they contain an exponent of $\eta_2+\eta_3$ which is conjugate to a pinned field and hence its correlation function decays exponentially.
We thus look for an operator that couples to the superconductor but does not contain $\eta_2+\eta_3$. The leading operator of this form is
\bea
\label{eq:gDelta}
O_n^\Delta = g_\Delta \psi_1^\dagger (\psi_1^\dagger {\psi_2})^{n} ({\psi_3} \psi_4^\dagger)^{n} \psi_4^\dagger
 \sim g_\Delta \cos (\eta_1 + \eta_4).
\eea
(In the wire the operator $\propto \cos(\eta_1-\eta_4)$ does not conserve momentum and hence is absent in a translationally invariant system.)
This process is shown in Fig.~\ref{fg:3}b. For $n=1$ (at filling factor $\nu=1/3$) it is generated at first order in $\Delta$ and in the interaction, $g_\Delta \propto \Delta  U^2_{2k_F}$.  Thus we see that when the proximity coupling is strong enough this term will pin $\eta_1+\eta_4$ and hence the low energy theory will become gapped. A similar many body process will also generate a term $\cos(\eta_2 +\eta_3)$ which competes with the $g_B$ term, see Fig.~\ref{fg:3}a. As long as the latter is stronger, however, we may neglect this term.

We now claim that the resulting gapped phase has  fractional Majorana modes at the ends of the wire. To treat the boundaries of the finite length (L) wire we use the usual unfolding transformation~\cite{Giamarchi} which results in a boundary condition
\bea
\phi_{1}(x)=\phi_{2}(x),\;\phi_{3}(x)=\phi_{4}(x),\;(x=0,L).
\eea
In the new variables the boundary conditions become
\bea
\eta_{1}(x)=\eta_{2}(x),\;\eta_{3}(x)=\eta_{4}(x),\;(x=0,L).
\eea

\begin{figure}[h]
\begin{center}
\includegraphics*[width=70mm]{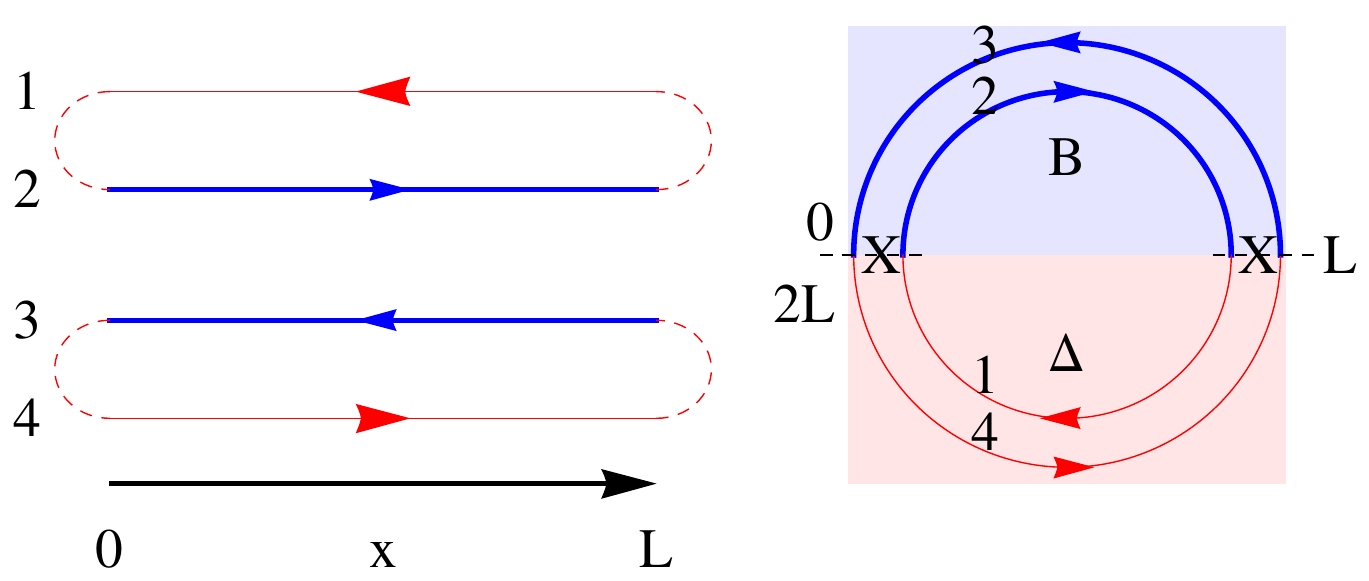}
\caption{\label{fg:4} Unfolding transformation on the spinfull finite wire to a ring whose upper part is coupled to the magnetic perturbation and whose lower part is coupled to the superconducting perturbation. Fractional Majorana modes form at $x=0$ and $x=L$.
}
\end{center}
\end{figure}

It proves convenient to define a theory with a single left moving field $\xi_L(x)$ and a single right moving field $\xi_R(x)$ on a ring of circumference $2L$, according to
\bea
\xi_{R(L)}(x)=\begin{cases}
   \eta_{2(3)}(x),  & 0 \le x \le L,\\
   \eta_{1(4)}(2L-x), & L \le x \le 2L,
 \end{cases}
\eea
see Fig.~(\ref{fg:4}).
The magnetic and superconducting perturbations to the Hamiltonians now act in different regions,
\bea
\delta \mathcal{L} =  \int_0^{2L} dx g_B(x) \cos (\xi_R-\xi_L)+g_\Delta(x) \cos( \xi_R+\xi_L),
\eea
where $g_B$ is finite in the region $ 0 \le x \le L$ and $g_\Delta$ is finite in the region $L \le x \le 2L$. For $n=0$ the interface between two gapped regions dominated by $\cos(\xi_R-\xi_L)$ and $\cos(\xi_R+\xi_L)$, binds Majorana zero modes. For $n\ne 0$, the interfaces bind fractional Majorana modes as was recently demonstrated in~\cite{Clarke12,Lindner12,Cheng12,Vaezi13}.

In principle, such zero modes lead to a degeneracy of the ground state. However, this degeneracy is apparent only when the system hosts at least four zero modes, such that there are several ground states that share the same fermion parity. Systems that host four zero modes may be double wire systems, or a single wire divided to segments of phases that are topologically different. In the present context, we may imagine constructing such segments by having regions where the $g_B,g_\Delta$ are operative, alternating between regions where $g_\Delta$ is put to zero, and the $\eta_1,\eta_4$ modes are gapped by a perturbation of the form
\begin{equation}
g_{B'} \int_a^b \cos\left(\eta_1-\eta_4\right) dx
\label{eq:gbprime}
\end{equation}
where $a,b$ are the end-points of the segment. Physically this term corresponds to a spatially oscillating magnetic field with wave vector $4 k_{SO}$. Alternatively, four zero modes may be constructed in systems of two wires with properly-tuned couplings that allow also for a transfer of charge between the wires.

   Disorder has a destabilizing influence on the gapped superconducting phase and its degeneracy. Indeed, a backscattering term  of the form $\cos{\left [\alpha(\eta_1-\eta_4)+\beta(\eta_2-\eta_3)\right ]}$ will lift the degeneracy, as now for example the states with $\eta_3-\eta_2=0$ and $\eta_3-\eta_2= 2\pi$ have different energies.

\emph{Relevance and irrelevance:} The fractional helical states, as well as the superconducting phases, can be stabilized if both $g_B$ and $g_\Delta$ are large enough. But yet a relevant question is under which conditions it is possible to trigger those instabilities even when the bare perturbations $g_B$ and $g_\Delta$ are small. We will use the renormalization group (RG) method to address this question. We note, however, that for finite one-dimensional wires, the RG flow is cut-off by the wire's length, and does not necessarily get to its asymptotic value. Therefore, the relevance of perturbations to actual experimental systems will strongly depend on the bare values of couplings.

In an interacting Luttinger liquid described by charge and spin interaction parameters $K_{c}$ and $K_s$, respectively, we find that the operators $\mathcal{O}_n^B$ [in Eq.~(\ref{eq:O})] and $O_n^\Delta$ [in Eq.~(\ref{eq:gDelta})] have scaling dimension
\begin{eqnarray}
x_{B}&=&\frac{1}{2}(K_{s}^{-1}+(2n+1)^2 K_{c}), \nonumber \\
x_{\Delta}&=&\frac{1}{2}(K_{c}^{-1}+(2n+1)^2 K_{s}),
 \end{eqnarray}
respectively. For small magnetic field we have $K_s=1$ due to an SU(2) symmetry of model~(\ref{eq:H}) realized in terms of fermions $\psi'_l(x) =e^{i s_l x k_{SO}} \psi_l(x)$ with $s_1=s_2=-s_3=-s_4=1$~\cite{Stoudenmire11}.

Consider first the magnetic perturbation $g_B$. If the density is fixed at $\nu=1/(2n+1)$, then the instability into the fractional helical state occurs for infinitesimal magnetic field when the scaling dimension is equal or smaller than the space-time dimension, namely at $K_c \le 3/(2n+1)^2$. If $k_{\rm F}$ deviates from the special value $k_{\rm SO}/(2n+1)$ then a non zero bare value of $g_B$ is essential to stabilize the helical phase through a commensurate-incommensurate (C-IC) transition. As known from the study of C-IC transitions~\cite{Giamarchi} at the transition the dimension $x_B$ takes the value one. Repulsive interactions reduce the value of the critical (bare) $g_B$ \cite{Sela11a}.

 A rough estimate of $K_c$ in wires shows that it may actually be rather small. To linear order in the interaction $U_q$ the charge Luttinger parameter is given by $K_c=\left(1+ \frac{4 U_q }{\pi \hbar v_F} \right)^{-1/2}$.
 Since the Coulomb interaction $U_q$ diverges logarithmically with the distance to the gates and near the band bottom the density of states $1/v_F$  diverges as well, their combination leads to a small $K_c$.

The superconducting perturbation $g_\Delta$ does not oscillate in space and hence does not require a special density. However, except for $n=0$, it is irrelevant, $x_\Delta > 2$. Indeed, even in the $n=0$ case it was pointed out that the superconducting gap is reduced due to interactions~\cite{Gangadharaiah11}. Thus, strong proximity coupling is necessary.

We note that apart from the couplings $g_B$ and $g_\Delta$, the system may become unstable with respect to spin density wave formation $\mathcal{O}_{SDW} = \psi^\dagger_{1} \psi_{2} \psi_{3} \psi_{4}^\dagger$. Note that there is no special filling condition for this operator to conserve momentum and this operator is exactly marginal for all values of $K_c$. The strong coupling picture of this instability corresponds to an antiferromagnetic spin arrangement with wave vector $2 k_F$.
This instability can be inhibited by applying a magnetic field  parallel to the $\hat y$ axis, shifting the up-spin and down-spin bands vertically. When that happens the SDW term does not conserve momentum since at the chemical potential the spin up Fermi momentum $k_{{\rm F}\uparrow}= (k_2 - k_1)/2$ is not equal to the spin down Fermi momentum $k_{{\rm F}\downarrow}= (k_4 - k_4)/2$ and the corresponding SDW phase can not be stabilized.

\emph{Summary:} In this manuscript we showed theoretically  that clean spinfull wires with spin-orbit coupling and strong electron-electron interaction may give rise to novel fractional helical liquids. The two terminal conductance of these liquids is quantized to a fractional value. We argued that in proximity to a superconductor fractional Majorana modes may be stabilized. We discussed the influence of disorder on these wires and pointed out that alternative systems such as the two coupled wires (that do not have substantial spin-orbit effects) may give rise to a similar fractionalization.

We expect that the ideas put forward here for fractionalized one-dimensional systems coupled to a superconductor can be extended, by studying several wires connected in parallel, to novel two-dimensional fractionalized superconducting phases. Furthermore,  it was shown in Ref.~\onlinecite{Teo11}
that two dimensional non-abelian quantum Hall systems whose quasi-particles allow for universal topological quantum computation may be viewed as constructed out of an array of parallel wires, with properly tuned inter-wire tunneling and interactions. We envision that a combination of such interaction, together with spin-orbit coupling and proximity coupling to a super-conductor may allow for such computation to be carried out in clean quasi-1D systems as well made of several wires.

We acknowledge discussions with J. Alicea, O. Auslaender,  M. Barkeshli, E. Berg,  D. Clarke,   L. Fidkowski, M. Freedman, N. Lindner, and A. Yacoby.
The work was supported by a BSF, ISF, DFG, Minerva  and Microsoft  grants.

\section{Appendix}
\subsection{Generalization to $N$ parallel wires}
\label{appendix:effectiveaction}
Kane \emph{et. al.}~\cite{Kane02} studied the 2D FQH system as many parallel spinless wires where a term of the form Eq.~(\ref{eq:O}) couples each pair of neighboring wires. We will now focus on such a system with a finite number of wires, $N$. This analysis will bridge between our system of interest, of a spinfull wire with SO coupling ($N=2$) and the 2D limit ($N \to \infty$).

Each wire $j=1,...,N$ is described by left ($\phi_{2j-1}$) and right ($\phi_{2j}$) boson fields. The chiral fermion operators are given by $\psi_{j}(x) \propto e^{i \phi_{j}(x)}$. The Lagrangian under consideration is $\mathcal{L}=\mathcal{L}_0+\delta \mathcal{L}$, where we assume that $\tilde U_{lj}$ of Eq.~(\ref{transformedL}) is diagonal and
\bea
\label{L0}
\mathcal{L}_0 =\frac{-1}{4 \pi}\sum_{j=1}^{2N} \left((-1)^j \partial_t \phi_{j} \partial_x \phi_{j} +v_F (\partial_x \phi_{j} )^2 + 2 E \phi_{j} (-1)^j \right).\nonumber \\
\eea
By a straightforward generalization of Eq.~(\ref{eq:transformation}) we define $\eta_j$ fields, $j=1,...,2N$, in terms of which the perturbation $\delta \mathcal{L}$  reads
\bea
\delta \mathcal{L}=\sum_{j=1}^{N-1} g_B  \cos\left(\eta_{2j} - \eta_{2j+1}\right).
\eea
 The fields $\eta_1$ and $\eta_{2N}$ will form the low energy sector providing the edge theory.

\subsection{Two-terminal conductance}
We now consider the two-terminal conductance in the case that the interactions act  in a finite region $-L/2 \le x \le L/2$. The analysis below is a direct generalizations of Eqs.~(\ref{eq:GS}), (\ref{bc1}), and (\ref{bc2}). Consider a situation where the non-interacting lead to the right has chemical potential $V$, and the lead to the left is at chemical potential $0$. We now describe the scattering problem in terms of incoming fields $I_i$ ($i=1,...,N$) and outgoing fields $O_i$ $(i=1,...,N)$. The conductance is given by
\bea
\label{GS}
G_N V &=& \sum_{l=1}^{N} (I_{2l-1}-O_{2l}) \nonumber \\&=& N V - (0,1,0,1,...)S(V,0,V,0,...) \nonumber \\
&=&\sum_{l=1}^{N} (-I_{2l}+O_{2l-1}) \nonumber \\&=& (1,0,1,0,...)S(V,0,V,0,...).
\eea
The $2N \times 2N$ scattering matrix $\vec{O} = S \vec{I}$ needs to be determined by a set of $2N$ equations. Generalizing the procedure in the main text, we get $(2N-2)$ equations by setting $\eta_{2i}=\eta_{2i+1}$ at $x=-L/2$ and at $x=L/2$,
\bea
(n+1) I_{2i} - n O_{2i -1} &=& (n+1) O_{2i+1} -n I_{2i+2}, \nonumber \\
(n+1) O_{2i} - n I_{2i -1} &=& (n+1) I_{2i+1} -n O_{2i+2},
\eea
where $i=1,2,...,N-1$. Two additional equations originate from the free propagation of $\eta_1$ and $\eta_{2N}$,
\bea
(n+1) I_1 - n O_2 &=& (n+1) O_1 -n I_2,\nonumber \\
(n+1) I_{2N} - n O_{2N-1} &=& (n+1) O_{2N} -n I_{2N-1}.
\eea
This set of equations can be written in matrix form $A \vec{O} = B \vec{I}$, giving $S=A^{-1} B$, and the two terminal conductance is found using Eq.~(\ref{GS}).

\begin{figure}[h]
\begin{center}
\includegraphics*[width=60mm]{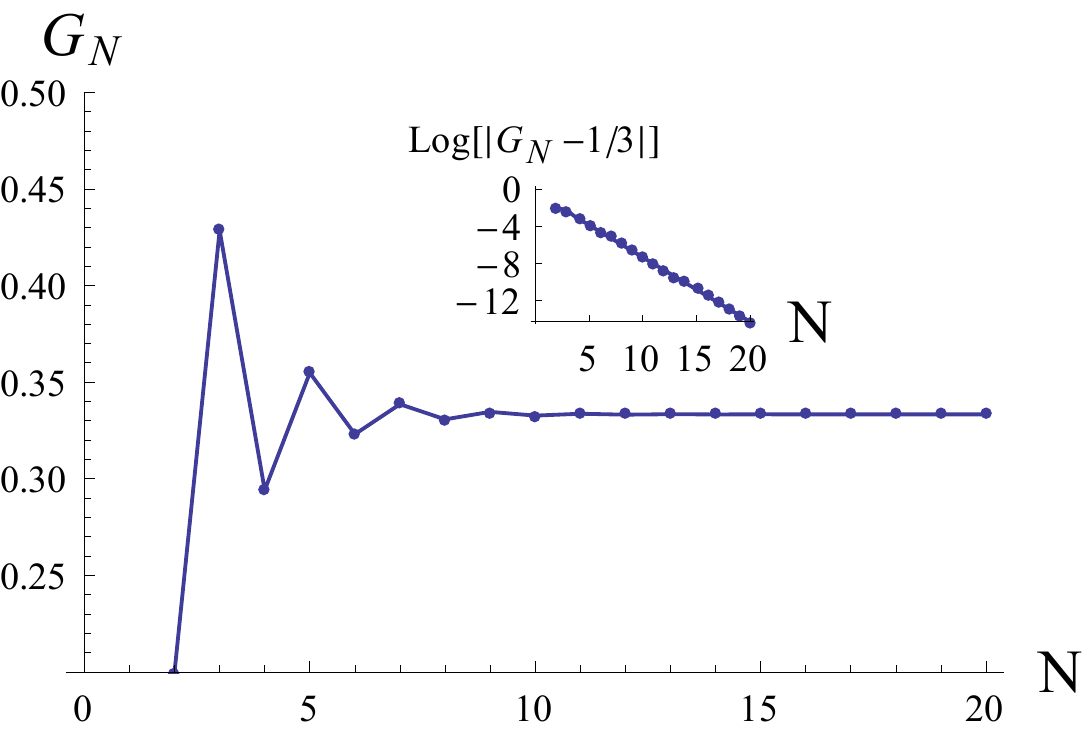}
\caption{\label{fg:Kn} Two terminal conductance on an array of $N$ wires for $n=1$ ($\nu=1/3$).
}
\end{center}
\end{figure}

The result of this calculation is that the conductance exponentially tends to $\frac{e^2}{h (2n+1)}$. Explicitly, we find that the result is given by Eq.~(\ref{eq:GN}) for all $\nu$ and $N$, which is plotted in Fig.~\ref{fg:Kn} for $n=1$ ($\nu=1/3$).

We remark that the sum of each column of the $S$ matrix is equal to unity, implying current conservation. However, $S$ can have negative elements. For example, for $N=2$,
\bea
S=\frac{1}{5}\left(
  \begin{array}{cccc}
    4 & 2 & -2 & 1 \\
    2 & 1 & 4 & -2 \\
    -2 & 4 & 1 & 2 \\
    1 & -2 & 2 & 4 \\
  \end{array}
\right).
\eea
For an incoming pulse (soliton) at $I_1$, we have as out going currents $\vec{O}=\frac{1}{5}(4,2,-2,1)$. The negative current in $O_3$ is not forbidden as it corresponds to an antisoliton.
\subsection{Low energy action}
We now derive the low energy effective action. This derivation will lead explicitly to the two freely propagating chiral modes as specific linear combinations of $\eta_1$ and $\eta_{2N}$, see Eq.~(\ref{phiRL}). For finite $N$ this effective theory is different from the edge theory of a FQH state, as dictated by the parameter $K_N$ derived below, determining the commutation relations of the chiral fields.

In terms of the $\eta$ fields the free Lagrangian becomes
\bea
\label{eq:L0}
\mathcal{L}_0 &=&-\sum_{j=1}^{2N}  [ (-1)^j\frac{\nu}{4 \pi} \partial_t \eta_j \partial_x \eta_j + \frac{E \nu}{2 \pi} \eta_j (-1)^j \nonumber \\ &+&\frac{v_F C}{8 \pi}  (\partial_x \eta_j)^2  +\frac{v_F D}{4 \pi}  \partial_x \eta_{2j-1} \partial_x \eta_{2j}  ],
\eea
with $C(D) = 1 \pm \nu^2$. We see that for $\nu \ne 1$ there is coupling between fields $ \eta_{2j-1}$ and $\eta_{2j} $ that will mediate a coupling between $\eta_1$ and $\eta_{2N}$. The effective action for $g_B \to \infty$ is obtained after integrating out the fields $\{ (\eta_{2j}+\eta_{2j+1})$, $j=1,...,N-1\}$ and setting $\eta_{2j}=\eta_{2j+1}$ ($j=1,...,N-1$).
Writing Eq.~(\ref{eq:L0}) in terms of $\eta_{2j}+\eta_{2j+1}$, and setting $\eta_{2j} -\eta_{2j+1}=0$ we find
\bea
\label{completesquare}
\mathcal{L}_0 &=&-\sum_{j=1,j=2N}  \big[ (-1)^j\frac{\nu}{4 \pi} \partial_t \eta_j \partial_x \eta_j + \frac{E \nu}{2 \pi} \eta_j (-1)^j \nonumber \\ &+&\frac{v_F C}{8 \pi}  (\partial_x \eta_j)^2 \big]\nonumber\\
&-&\frac{v_F
}{16 \pi} \sum_{l,j=1}^{N-1} \partial_x (\eta_{2l}+\eta_{2l+1}) \mathcal{U}_{lj}  \partial_x (\eta_{2j}+\eta_{2j+1}) \nonumber \\
&-& \frac{v_F  D}{8 \pi} \left( \partial_x (\eta_{2}+\eta_3) \partial_x \eta_1 + \partial_x (\eta_{2N-2}+\eta_{2N-1}) \partial_x \eta_{2N} \right) ,\nonumber
\eea
where the $\left(N-1\right) \times \left(N-1\right)$ matrix $\mathcal{U}$ is given by
\bea
\mathcal{U} = \left(
            \begin{array}{cccc}
              C & D/2 & 0 & \hdots \\
              D/2 & C & D/2 & ~ \\
              0 & D/2 & C & ~ \\
              \vdots & ~ & ~ & \ddots \\
            \end{array}
          \right).
\eea
 To get the interaction between $\eta_1$ and $\eta_{2N}$ we complete to a square the terms containing $\partial_x (\eta_2+\eta_3)$ and $\partial_x (\eta_{2 N-2}+\eta_{2N-1})$ and obtain the effective Lagrangian
\bea
&&\mathcal{L}_{\rm eff}[\eta_1,\eta_{2N}] =  -\frac{\nu}{4 \pi} (\partial_t \eta_{2N} \partial_x \eta_{2N} -\partial_t \eta_{1} \partial_x \eta_{1})\nonumber \\
&-& \frac{E \nu}{2 \pi} (\eta_{2N}-\eta_1)+ \frac{v_F}{16 \pi}\big[ (C+F_1)  (\partial_x (\eta_{1}+\eta_{2N}))^2 \nonumber \\
 &+&(C+F_2)   ( \partial_x (\eta_{1}-\eta_{2N}))^2\big],
\eea
with
\bea
F_{1,2} &=& - \left( \frac{D}{2} \right)^2   \\
&\times& \left(\mathcal{U}^{-1}_{1,1} \pm \mathcal{U}^{-1}_{N-1,1} \pm \mathcal{U}^{-1}_{1,N-1} + \mathcal{U}^{-1}_{N-1,N-1}  \right). \nonumber
\eea
Finally, to bring the action to a non-interacting form, we define new propagating eigenmodes
\bea
\label{phiRL}
2\tilde{\phi}_R&=&\eta_{2N}\left(\frac{1}{K_N}+\nu \right)+\eta_{1}\left(\frac{1}{K_N}-\nu \right),\nonumber \\
2\tilde{\phi}_L&=&\eta_{1}\left(\frac{1}{K_N}+\nu \right)+\eta_{2N}\left(\frac{1}{K_N}-\nu \right),
\eea
and obtain
\bea
\mathcal{L}_{{\rm{eff}}} &=&-\frac{1}{4 \pi} \\
&\times&\sum_{p} \left(K_N p \partial_t \tilde{\phi}_{p} \partial_x \tilde{\phi}_{p} +\tilde{u} K_N (\partial_x \tilde{\phi}_p )^2 + 2 p E \tilde{\phi}_{p}  \right).\nonumber
\label{effL}
\eea
Here $p=R/L=\pm 1$. $K_N$ and $\tilde{u}$ are given by
\bea
\frac{1}{K_N} =\nu \sqrt{\frac{C+ F_1}{C+ F_2}},~~\tilde{u}=\pi \nu^{-1} \sqrt{(C+F_1)(C+F_2)}.
\eea
 Here $\tilde{u}$ is identified as the velocity. For $N=2$ the matrix $\mathcal{U}$ has one element, $C$, and one obtains $K_2 = \frac{2 }{\nu^{-2}+1}$. Generally we find that this definition of $K_N$ coincides with the result for the two terminal conductance $K_N^{-1}=G_N/(e^2/h)$ [where the general equation for $G_N$ is given by Eq.~(\ref{eq:GN})]. Thus, also $K_N^{-1}$ tends exponentially to the value $\nu$, where the effective theory reduces to two independent chiral FQH edge theories~\cite{Wen90a}. The total density is $\rho =  \frac{1}{2 \pi} \partial_x \tilde{\phi}_R -  \frac{1}{2 \pi} \partial_x \tilde{\phi}_L$. In the DC limit variation with respect to $\partial_x \tilde{\phi}_p$ gives the compressibility is $\frac{d \rho}{d \mu} = \frac{1}{\pi \tilde{u} K_N}$.

 Perturbations to the helical liquid phase of the quantum wire, such as impurity scattering,  can be studied via renormalization group analysis with respect to the fixed point of  Eq.~(\ref{effL}) (and setting $N=2$ for the case of a quantum wire).  The scaling dimension of the quasiparticle operators $ e^{i \tilde{\phi}_{p}}$  is $\frac{1}{2 K_N}$ ($p=L/R$).
 Using Eq.~(\ref{phiRL}) we see that
 \bea
 \nu(\eta_1 - \eta_{2N}) = \tilde{\phi}_L -  \tilde{\phi}_R.
 \eea
 Hence the dimension of $\cos \left[ \alpha (\eta_1 - \eta_{2 N}) \right]$ is
  \bea
  \label{scalingD}
  \frac{\alpha^2}{\nu^2 K_N},
  \eea determining the scaling dimension of the various impurity perturbations discussed in the text.


\end{document}